\documentclass[usenatbib]{mn2e}
\pdfoutput=1
\usepackage[pdftex]{graphicx}
\def\twid{\mathrel{\lower.1ex\hbox{$\sim$}}}
\def\gtwid{\mathrel{\raise.3ex\hbox{$>$\kern-.75em\lower1ex\hbox{$\sim$}}}}
\def\ltwid{\mathrel{\raise.3ex\hbox{$<$\kern-.75em\lower1ex\hbox{$\sim$}}}}
\def\\{\hfil\break}

\newcommand{\be}{\begin{equation}}
\newcommand{\ee}{\end{equation}}
\newcommand{\bea}{\begin{eqnarray}}
\newcommand{\eea}{\end{eqnarray}}

\begin{document}

\title{The effect of massive neutrinos on the matter power spectrum}
\author[Agarwal \& Feldman]{
Shankar Agarwal$^{1, \dagger}$\,\& Hume A. Feldman$^{1,\star}$\\
$^1$Department of Physics \& Astronomy, University of Kansas, Lawrence, KS 66045, USA.\\
emails: $^{\dagger}$sagarwal@ku.edu; $^{\star}$feldman@ku.edu}
\date{} 

\maketitle

\begin{abstract}
We investigate the impact of massive neutrinos on the distribution of matter in the semi-non-linear regime $(0.1\!\ltwid\!k\!\ltwid\!0.6\, h \textrm{Mpc}^{-1})$. We present a suite of large-scale $\it{N}$-body simulations quantifying the scale dependent suppression of the total matter power spectrum, resulting from the free-streaming of massive neutrinos out of high-density regions. Our simulations show a power suppression of $3.5-90$ per cent at $k\!\twid\!0.6\,h\textrm{Mpc}^{-1}$ for total neutrino mass, $\Sigma m_\nu=0.05-1.9\,{\rm eV}$ respectively. We also discuss the precision levels that future cosmological datasets would have to achieve in order to distinguish the normal and inverted neutrino mass hierarchies.
\end{abstract}

\noindent{\it Subject headings}: neutrinos -- methods: numerical -- large-scale structure of Universe.



\section{Introduction}
In the standard model of particle physics there are three types (flavours) of neutrinos: electron neutrino ($\nu_e$), muon neutrino ($\nu_\mu$) and tau neutrino ($\nu_\tau$). Neutrino oscillation experiments \citep{KamLAND, SNO} in the past decade indicate that at least two neutrino eigentstates have non-zero masses. The direct implication of massive neutrinos is a non-zero hot dark matter (HDM) contribution to the total energy density of the Universe. Being sensitive to the mass squared differences between the neutrino eigentstates, the oscillation experiments only provide a lower bound on the total neutrino mass. Mass splittings of $|\Delta m^{2}_{32}|=(2.43\pm0.13)\times10^{-3}\,{\rm eV}^2$ and $\Delta m^{2}_{21}=(7.59\pm0.21)\times10^{-5}\,{\rm eV}^2$ \citep{MINOS,KamLAND} imply a lower limit for the sum of the neutrino masses to be $0.05$ and $0.1\,{\rm eV}$ for the normal and inverted mass hierarchies \citep{OttenWein08}, respectively.

During the radiation era, matter perturbations on the sub-horizon scales grow logarithmically. The earlier a mode enters the horizon, the more it is suppressed due to the decaying gravitational potentials. On the other hand, the superhorizon modes do not decay until they enter the horizon. As a result, the matter power spectrum turns over at a scale that corresponds to the one that entered the horizon at radiation--matter equality. Neutrinos with mass on the sub-eV scale behave as a hot component of the dark matter. Neutrinos stream out of high-density regions into low-density regions, thereby damping out small-scale density perturbations. Massive neutrinos, therefore, suppress the logarithmic growth of sub-horizon modes. Extremely low mass neutrinos become non-relativistic after the radiation era is over and the free-streaming damping of matter perturbations affects even those scales that were always outside the horizon during the radiation era.

The redshift-dependent free-streaming comoving wave number, $k_{\rm fs}$, is given by
\be
k_{\rm fs}(z)=\sqrt{\frac{3}{2}}\frac{H(z)}{v_{\rm th}(1+z)},
\ee
where $H(z)$ and $v_{\rm th}$ are the Hubble parameter and the neutrino thermal velocity, respectively. As long as neutrinos are relativistic, they travel at the speed of light and their free-streaming comoving wave number shrinks at the same rate as that of the comoving Hubble wave number (equation 1). After a neutrino eigentstate becomes non-relativistic, its thermal velocity decays as
\bea
v_{\rm th}&\approx&\frac{3T_\nu}{m_\nu}=3\left(\frac{4}{11}\right)^{1/3}\frac{T^0_\gamma(1+z)}{m_\nu}\nonumber\\
&\approx&151(1+z)\left(\frac{1\,{\rm eV}}{m_\nu}\right) \textrm{km/s},
\eea
where $m_\nu$ is the mass of a neutrino eigentstate in eV and the present-day photon temperature, $T^0_\gamma$, is 2.725 K \citep{WMAP7a}.

Thus the free-streaming comoving wave number for non-relativistic neutrinos is given by
\be
k_{\rm fs}\approx0.81\frac{\sqrt{\Omega_\Lambda+\Omega_{\rm m}(1+z)^3}}{(1+z)^2}\left(\frac{m_\nu}{1\,{\rm eV}}\right)h\, \textrm{Mpc}^{-1}.
\ee
For a massive eigentstate, the redshift of non-relativistic transition ($m_\nu\approx3T_\nu$) is given by
\be
1+z_{\rm nr}\approx1987\left(\frac{m_\nu}{1\,{\rm eV}}\right).
\ee
After a neutrino eigentstate becomes non-relativistic, $k_{\rm fs}$ begins to grow as $k_{\rm fs}\propto(1+z)^{-1/2}$. Thus, $k_{\rm fs}$ passes through a minimum, $k_{\rm nr}$, which can be shown to be (from equation 3)
\be
k_{\rm nr}\approx0.018\left(\frac{m_\nu}{1\,{\rm eV}}\right)^{1/2}(\Omega_{\rm m}h^2)^{1/2}\, \textrm{Mpc}^{-1}.
\ee

For modes with $k>k_{\rm fs}$, the neutrino density perturbations are erased. This weakens the gravitational potential wells and the growth of cold dark matter (CDM) perturbations is suppressed. Perturbations are free to grow again once their comoving wave numbers fall below $k_{\rm fs}$. Modes with $k<k_{\rm nr}$ are never affected by free-streaming and neutrino perturbations evolve like CDM perturbations. Baryon density perturbations, on the other hand, being pressure supported, can grow in amplitude only after photon decoupling. At the time of photon decoupling, baryons fall into the neutrino-damped dark matter potential wells. Thus, accurate measurements of the amplitude of clustering of matter in the Universe can provide strong upper bounds on the mass of neutrinos.

Section 2 describes how we implement neutrinos in our $\it{N}$-body simulations. In this section we also discuss the numerical methods employed in some complementary recent studies. In section 3 we discuss the convergence tests for the matter power spectrum calculated from our $\it{N}$-body simulations. In section 4 we show the impact of massive neutrinos on the matter distribution through the total matter power spectrum. In section 5 we discuss, based on our $\it{N}$-body simulations, the precision levels at which future galaxy surveys would need to measure the matter power spectrum, in order to distinguish between the normal and inverted mass hierarchies. In section 6 we compare our results with the neutrino simulations performed by other groups. In section 7 we estimate the errors in our $\it{N}$-body matter power spectra. We present our conclusions in section 8.

\section{Implementing Neutrinos in the $\it{N}$-body Simulations}

Neutrinos in the mass range $0.05<\Sigma m_\nu<1\,{\rm eV}$ have present-day free-streaming scales $0.04<k_{\rm fs}< 0.3\, h \textrm{Mpc}^{-1}$ $(150>\lambda_{\rm fs}>20\, h^{-1} \textrm{Mpc})$ and thermal velocities $3000>v_{\rm th}>450\textrm{\,km/s}$ respectively. Such large thermal velocities would prevent neutrinos from clustering with CDM and baryons, thereby keeping the neutrino perturbations in the linear regime. As such, in our simulations, we have assumed that the non-linear neutrino perturbations can be ignored and include the linear neutrino perturbations in the initial conditions (ICs) only.\\

To generate the ICs for CDM particles and baryons, we use the publicly available {\sc camb} code \citep{CAMB} and {\sc enzo}1.5 code\footnote{http://lca.ucsd.edu/projects/enzo}  \citep{SheBryBorNor04, NorBryHarBorReySheWag07} -- an adaptive mesh refinement (AMR), grid-based hybrid code (hydro + $\it{N}$-Body) designed to simulate cosmological structure formation. We use the {\sc camb} code to calculate the linear transfer functions for a given CDM$+$baryon$+$neutrino$+\Lambda$ model. The linear density fluctuation field for CDM particles and baryons is then calculated from their transfer functions using {\sc enzo}1.5. The initial positions and velocities for CDM particles and baryon velocities are calculated using the Zel'dovich Approximation (ZA, \cite{ZA70}). Note that we do not have neutrinos in our simulations as $\it{N}$-body particles or as a linear grid. Neutrinos enter our simulations only as neutrino-weighted CDM and baryon transfer functions from {\sc camb}.

The linear matter power spectrum, $P^{\rm L}_{\rm m}$, can be calculated at $z=0$ as the weighted average of the neutrino ($P^{\rm L}_\nu$) and the combined CDM plus baryon($P^{\rm L}_{\rm cb}$) linear spectra:
\be
P^{\rm L}_{\rm m}(k)=\left(
	(f_{\rm c}+f_{\rm b})\sqrt{P^{L}_{\rm cb}(k)}+f_\nu\sqrt{P_\nu^{\rm L}(k)}
	\right)^2,
\ee
where the weights are $f_{\rm i}=\Omega_{\rm i}/\Omega_{\rm m}$ and $\Omega_{\rm m}=\Omega_{\rm b}+\Omega_{\rm c}+\Omega_\nu$. The CDM plus baryon power spectrum is
\be
P^{\rm L}_{\rm cb}(k)=(f_{\rm c}+f_{\rm b})^{-2}\left(f_{\rm c}\sqrt{P_{\rm c}^{\rm L}(k)} + f_{\rm b}\sqrt{P_{\rm b}^{\rm L}(k)}\right)^2,
\ee
where $P^{\rm L}_{\rm c}$ and $P^{\rm L}_{\rm b}$ are the linear CDM and baryon power spectra respectively. The superscript `L' indicates quantities in the linear regime. On smaller scales the matter perturbations have gone non-linear. So, the non-linear matter power spectrum, $P^{\rm NL}_{\rm m}$, at $z=0$ becomes
\be
P^{\rm NL}_{\rm m}(k)=\left(
	(f_{\rm c}+f_{\rm b})\sqrt{P^{\rm NL}_{\rm cb}(k)}+f_\nu\sqrt{P_\nu^{\rm L}(k)}
	\right)^2,
\ee
where,
\be
P^{\rm NL}_{\rm cb}(k)=(f_{\rm c}+f_{\rm b})^{-2}\left(f_{\rm c}\sqrt{P_{\rm c}^{\rm NL}(k)} + f_{\rm b}\sqrt{P_{\rm b}^{\rm NL}(k)}\right)^2.
\ee

In equation (8), we calculate $P^{\rm NL}_{\rm cb}$ at $z=0$ from $\it{N}$-body simulations and combine it with $P_\nu^{\rm L}$ at $z=0$ as solved by the {\sc camb} code to construct $P^{\rm NL}_{\rm m}(k)$. Note that we do not account for the non-linear neutrino corrections in equation (8). \cite{SaiTakTar09} studied the non-linear neutrino perturbations using the higher-order perturbation theory (PT) to show that for low neutrino fractions ($f_\nu\!\ltwid\!0.05$), the amplitude of the non-linear matter power spectrum increases by $\!\ltwid\!0.01$ per cent at  $k\!\twid\!0.2\, h \textrm{Mpc}^{-1}$ at $z=3$ and by $\!\ltwid\!0.15$ per cent at  $k\!\twid\!0.1\,h\textrm{Mpc}^{-1}$ at $z=0$. Since at $z=0$, the PT approach to the non-linear matter power spectrum is expected to reproduce the $\it{N}$-body simulation results within $1$ per cent -- only for $k\!\ltwid\!0.1-0.15\,h\textrm{Mpc}^{-1}$ \citep{TarNisSaiHir2009}, the non-linear neutrino corrections at $z=0$ may be somewhat larger on scales we probe in our simulations $(0.1  \leq k \leq 0.6\,h\textrm{Mpc}^{-1})$ -- the estimate of which requires multiple particle (CDM$+$baryon$+$neutrino) simulations.

Numerical studies of the effect of neutrinos on the matter distribution have recently been performed independently by \cite{BraHan08,BraHan09,BraHan10} and \cite{VieHaeSpr10}. Both groups choose similar cosmological parameters: ($\Omega_{\rm m}=0.3, \Omega_{\rm b}=0.05, \Omega_{\rm c}+\Omega_\nu=0.25, \Omega_\Lambda=0.7, h=0.7, n_{\rm s}=1)$, a $512\, h^{-1}\textrm{Mpc}$ box and an initial redshift for simulations, $z_{\rm i}=49$. \cite{BraHan08} and \cite{BraHan09,BraHan10} use a weighted sum of the CDM$+$baryon transfer functions (since they do not have baryons in their simulations) to generate ICs for the CDM component using ZA$+$second-order Lagrangian perturbation theory (2LPT; \cite{Sco98}). The \cite{VieHaeSpr10} simulations include baryons and use ZA to generate ICs. Both groups include neutrinos in their $\it{N}$-body simulations either as $\it{N}$-body particles, as a linear grid or use a hybrid method where neutrinos are treated as grid or particles depending on their thermal motion. In the grid-based implementation, the neutrino grid is evolved linearly and does not include the non-linear corrections. The particle-based implementation accounts for the non-linearities by including the coupling between the gravitational potential and neutrinos.

\cite{BraHan09} (their fig. 1, middle panel) show that the error from neglecting non-linear neutrino perturbations at $z=0$ is at most $1.25$ per cent level at $k\!\twid\!0.25\,h\textrm{Mpc}^{-1}$ for $\Sigma m_\nu=0.6\,{\rm eV}$. Also, the error between the grid and particle representations is shown to become smaller on small scales. Specifically, the two representations converge for $k\!\gtwid\!0.2\,h\textrm{Mpc}^{-1}$. This is attributed to the fact that the neutrino white noise (due to the finite number of neutrino $\it{N}$-body particles) contribution to the matter power spectrum dominates only on ever smaller scales as the CDM perturbations grow at low redshifts. \cite{VieHaeSpr10} (their fig. 2, right panel) show that the non-linear correction at $z=0$ may be as high as $6$ per cent at $k\!\twid\!1\,h\textrm{Mpc}^{-1}$ for $\Sigma m_\nu=0.6\,{\rm eV}$ and the agreement between the grid and particle representations begins to improve only at $k\!\gtwid\!1\,h\textrm{Mpc}^{-1}$. The discrepancies between the results from the two groups worsens significantly when the above comparison is done at $z=1$. These large discrepancies can not be solely due to the absence/presence of baryons or whether ZA or ZA$+$2LPT is used to generate the ICs since (i) the baryons closely trace the CDM distribution on scales $k\!\ltwid\!1\, h \textrm{Mpc}^{-1}$ and (ii) ZA or ZA$+$2LPT do not affect the final results significantly when the simulations start at a high redshift ($z_{\rm i}=49$). The extent and the scale-dependence of non-linear neutrino corrections are still being researched.

\section{$\it{N}$-body Simulations: Optimizing Boxsize and Number of Particles}
We performed $\it{N}$-body simulations with the {\sc enzo}1.5 code. The code allows us to choose the geometry (box size, number of particles), the cosmology ($\Omega_{\rm m}, \Omega_{\rm b}, \Omega_\Lambda, \Omega_\nu$), the amplitude of fluctuation on 8$\,h^{-1}$ Mpc scale: $\sigma_8$, the primordial spectral index: $n_{\rm s}$ and the initial redshift: $z_{\rm i}$. We kept AMR off (no adaptive mesh refinement) since it does not significantly affect the scales of interest. Throughout this paper we assume the 7-yr $\it{Wilkinson\, Microwave\, Anisotropy\, Probe}$ ($\it{WMAP}$; \cite{WMAP7b}) central parameters: $\Omega_{\rm m}=0.266$, $\Omega_{\rm b}=0.044$, $\Omega_\Lambda=0.734$, $h=0.71$ and  $n_{\rm s}=0.963$ for the matter, baryonic and cosmological constant normalized densities, the Hubble constant and the primordial spectral index respectively. We vary $\Omega_\nu$ such that $\Omega_{\rm cdm}+\Omega_\nu=0.222.$ The simulation parameters are listed in Table 1. In order to suppress sampling variance of the estimated power spectrum, for each row we ran eight simulations by changing the seed to generate the ICs.

\vspace{\baselineskip}
\begin{table}
\begin{center}
\begin{tabular}{| c | l | l | l | |}
\hline
Box size ($h^{-1}\textrm{Mpc}$)& $N_{\rm cdm}$ & $N_{\rm gas}$ & $\Omega_\nu$ \\ \hline
$200$ & $64^3$ & $512^3$ & $0.00$ \\
$200$ & $128^3$ & $512^3$ & $0.00$ \\
$200$ & $256^3$ & $512^3$ & $0.00$ \\
$200$ & $256 ^3$ & $512^3$ & $0.001$ \\
$200$ & $256 ^3$ & $512^3$ & $0.002$ \\
$200$ & $256 ^3$ & $512^3$ & $0.01$ \\
$200$ & $256 ^3$ & $512^3$ & $0.02$ \\
$200$ & $256 ^3$ & $512^3$ & $0.04$ \\
$100$ & $256 ^3$ & $512^3$ & $0.00$ \\
$200$ & $512^3$ & $512^3$ & $0.00$ \\
$200$ & $512 ^3$ & $512^3$ & $0.01$ \\
$200$ & $512 ^3$ & $512^3$ & $0.02$ \\
$200$ & $512 ^3$ & $512^3$ & $0.04$ \\
\hline
\end{tabular}
\end{center}
\caption{Simulation parameters. All simulations were started at a redshift of $z_{\rm i}=20$ and stopped at $z=0$. We ran eight independent simulations for each row to suppress sampling variance.}
\end{table}
\vspace{\baselineskip}

First, we had to select an appropriate geometry (box size and the number of CDM/gas particles) for which the matter power spectrum converges to $1$ per cent accuracy in the semi-non-linear regime $(0.1\!\ltwid\!k\!\ltwid\!0.6\, h \textrm{Mpc}^{-1})$. The largest mode that can fit in a $200\, h^{-1}\textrm{Mpc}$ box is $k\!\twid\!0.03\,h\textrm{Mpc}^{-1}$ and the matter power spectrum is sufficiently linear on these scales. One can choose bigger volumes but unless the number of particles is also increased accordingly, it leads to a poor mass resolution. Also, $\it{N}$-body simulations suffer from a discreteness problem that arises due to the finite number of macroparticles used to sample the matter distribution in the universe. Thus, given any theoretical cosmological model, the ICs are always undersampled.

The smallest scale for which the power spectrum can be resolved accurately is related to the Nyquist wavenumber, $k_{\rm Ny}$, given by:
\be
k_{\rm Ny}\!=\!\frac{\pi(N_{\rm part})^{1/3}}{L_{\rm Box}}
\ee
Given a combination of the number of particles and the box size, the power spectrum is dominated by shot noise for $k\!\gtwid\!k_{\rm Ny}$. For $N_{\rm cdm}\!=\!64^3$ particles in a $200\, h^{-1}\textrm{Mpc}$ box, $k_{\rm Ny}$ is $1.01\, h\textrm{Mpc}^{-1}$, while the semi-non-linear modes of interest are $0.1\!\ltwid\!k\!\ltwid\!0.6\, h \textrm{Mpc}^{-1}$. Thus $N_{\rm cdm}\!=\!64^3$ particles in a $200\, h^{-1}\textrm{Mpc}$ box seems a reasonable combination to start with.

The number of gas particles fixes the root grid that determines the force resolution for the simulation. {\sc enzo} uses a particle mesh technique to calculate the gravitational potential on the root grid \citep{SheNagSprHerNor03}. Forces are first computed on the mesh by finite-differencing the gravitational potential and then interpolated to the dark matter particle positions to update the particle's position and velocity information. This methodology requires that the root grid be at least twice as fine as the mean interparticle separation to obtain accurate forces down to the scale of the mean interparticle spacing. A coarse root grid renders the forces on the scale of the mean interparticle spacing, inaccurate. This explains our choice of $N_{\rm gas}\!=\!512^3.$

\begin{figure}
     \includegraphics[width= \columnwidth]{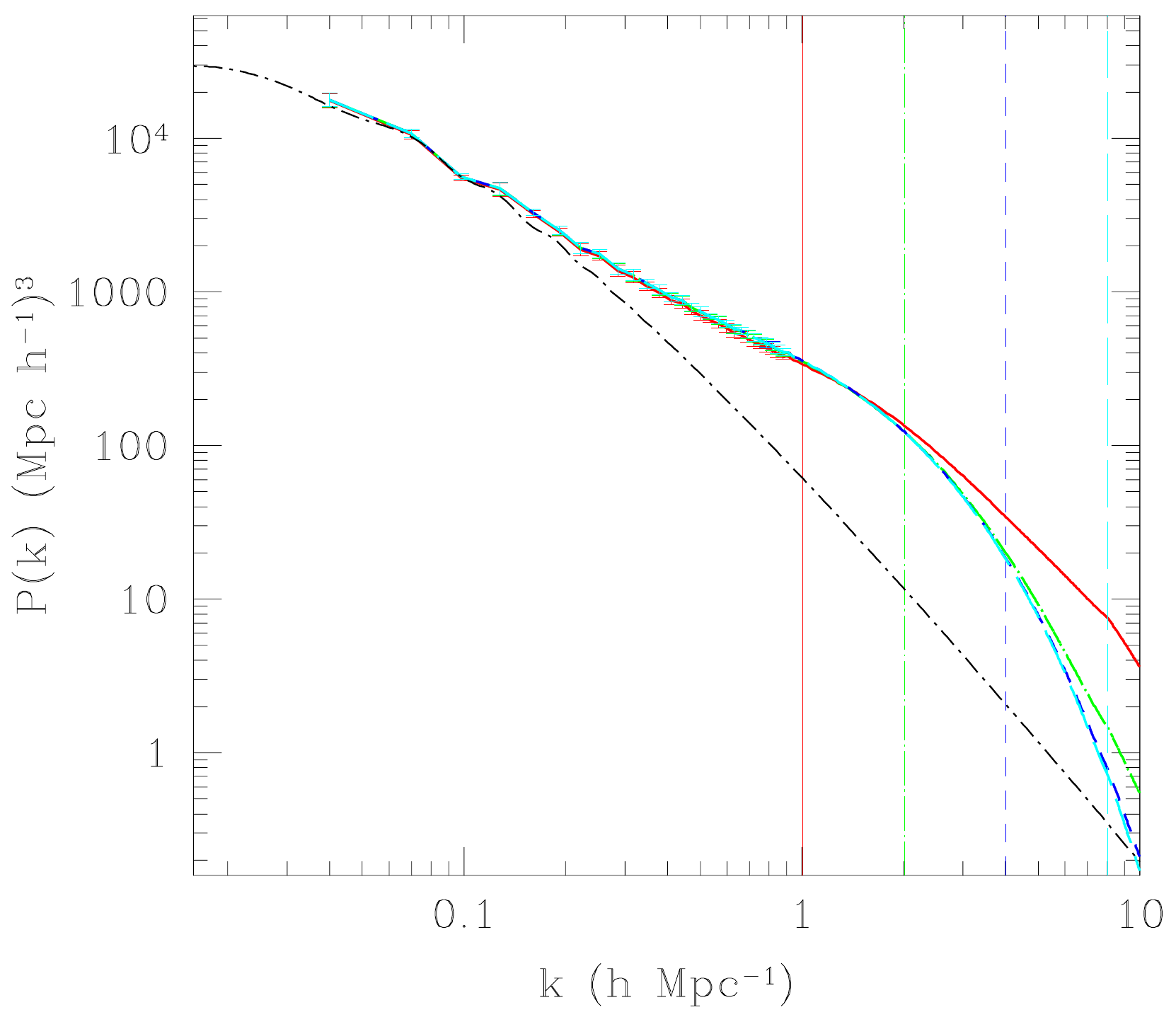}
        \caption{Matter power spectrum at $z=0$ for undersampled ICs at $z_{\rm i}\!=\!20$ with $N_{\rm cdm}\!=\!64^3- \textrm{solid (red)},\,128^3- \textrm{long dash--dotted (green)},\,256^3- \textrm{dashed (blue)}$ and $512^3- \textrm{long-dashed (cyan)}$. The vertical lines are the $k_{\rm Ny}$ wavenumbers for $64^3,\,128^3,\,256^3$ and $512^3$ CDM particles. Also plotted (dot--dashed line) is the linear theoretical power spectrum. For $k>k_{\rm Ny}$, particle shot noise dominates the true power spectrum.
        }
    \label{fig:undersampling1}
\end{figure}

\begin{figure}
     \includegraphics[width= \columnwidth]{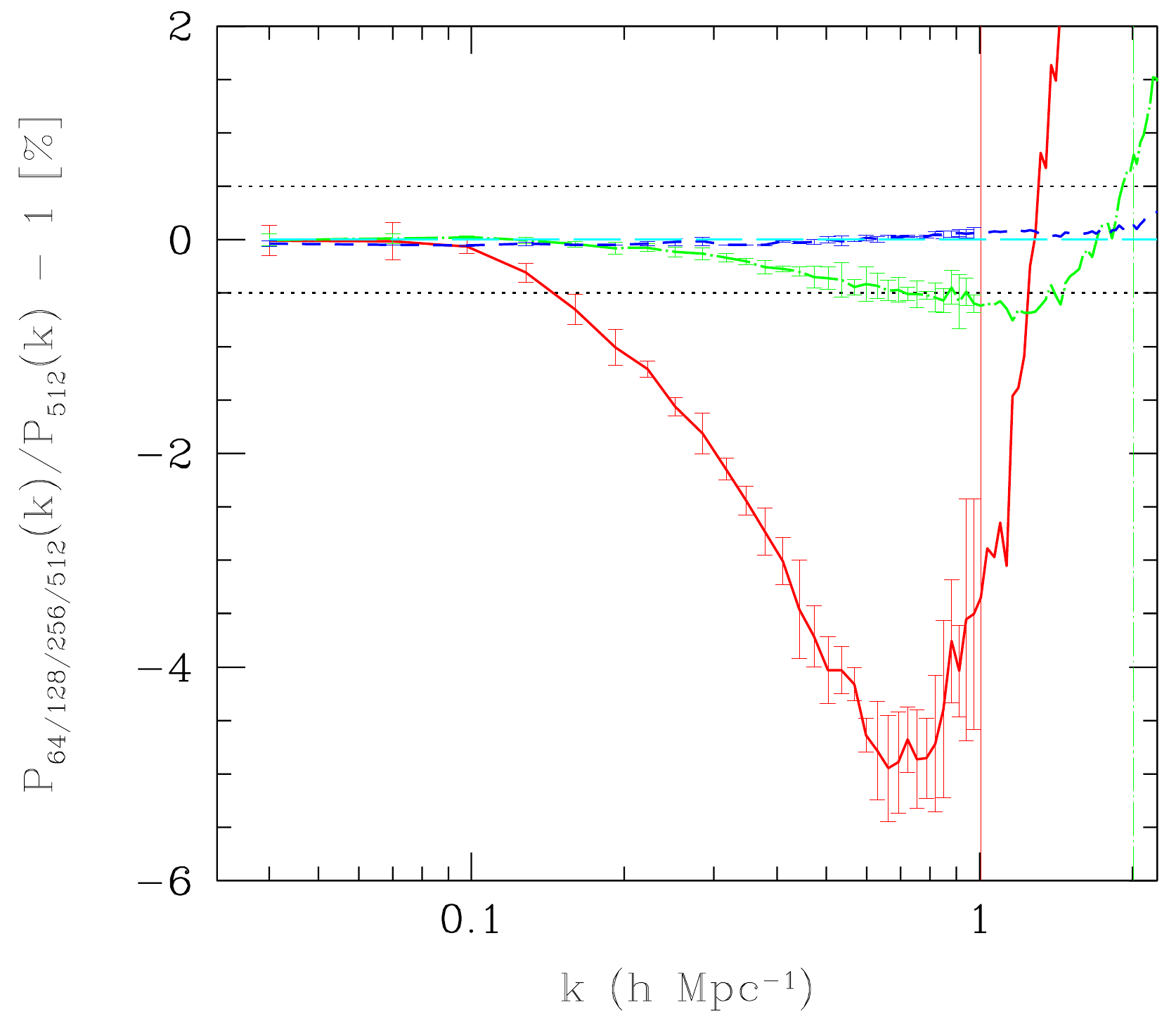}
        \caption{
Same as Fig.~\ref{fig:undersampling1} expressed as fractional suppression of the matter power spectrum at $z\!=\!0$ when $64^3- \textrm{solid (red)},\,128^3- \textrm{long dash--dotted (green)}$ and $256^3- \textrm{dashed (blue)}$ CDM particles are used to sample the ICs w.r.t the case where $512^3- \textrm{long-dashed (cyan)}$ CDM particles are used. $\Omega_\nu=0$ for all four cases.The error bars correspond to eight simulations with different seeds for the ICs.      
        }
    \label{fig:undersampling2}
\end{figure}

\begin{figure}
     \includegraphics[width= \columnwidth]{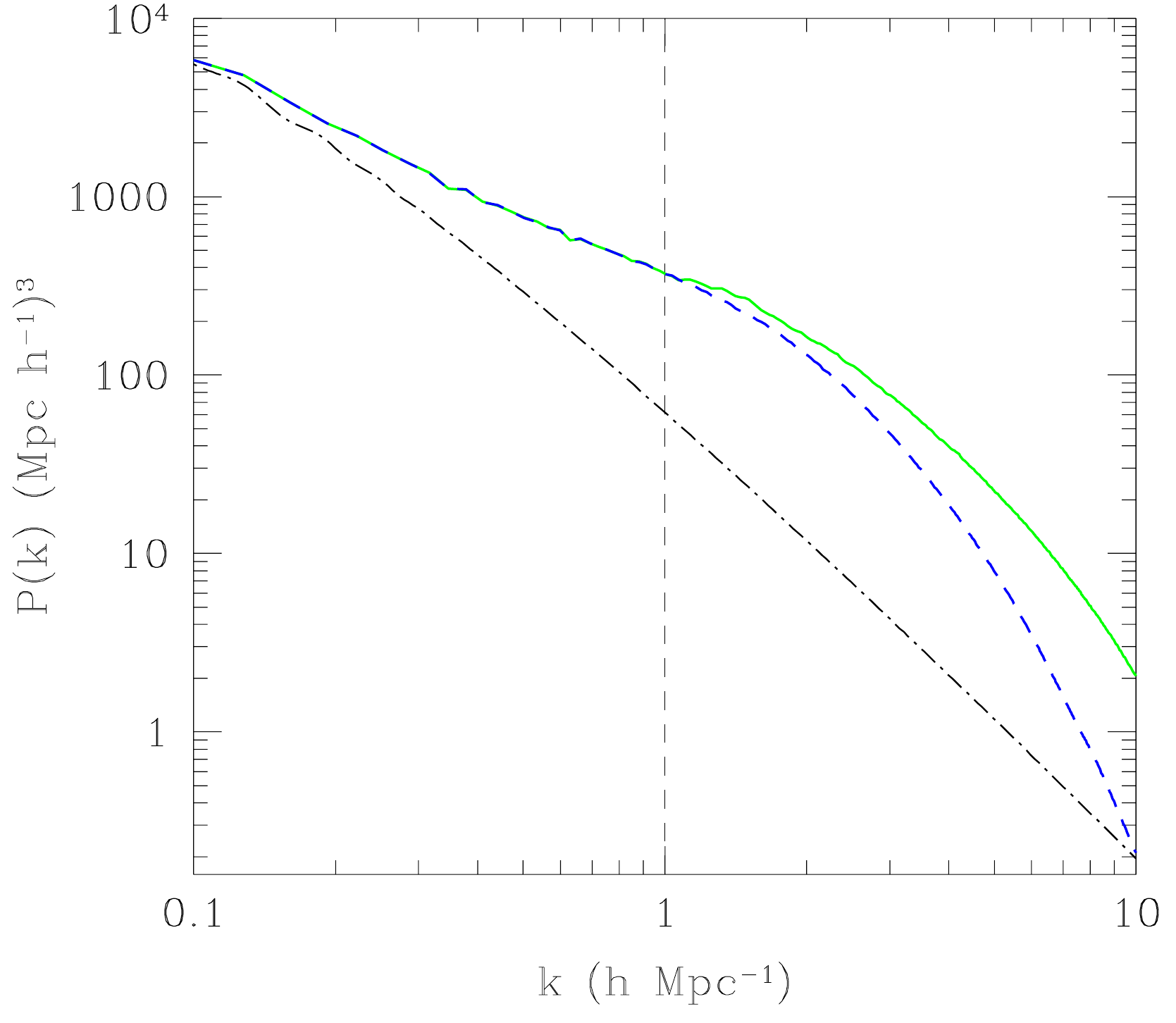}
        \caption{
Matter power spectrum at $z=0$ from $100\, h^{-1}\textrm{Mpc $-$ solid (green)}$ and $200\, h^{-1}\textrm{Mpc $-$ dashed (blue)}$ box simulations. The linear theory spectrum (dot--dashed line) is also shown. The vertical dashed line is the maximum wavenumber up to which the power spectrum from $200\, h^{-1}\textrm{Mpc}$ box simulations can be trusted at $1$ per cent level.
        }
    \label{fig:ps_box}
\end{figure}

\begin{figure}
     \includegraphics[width= \columnwidth]{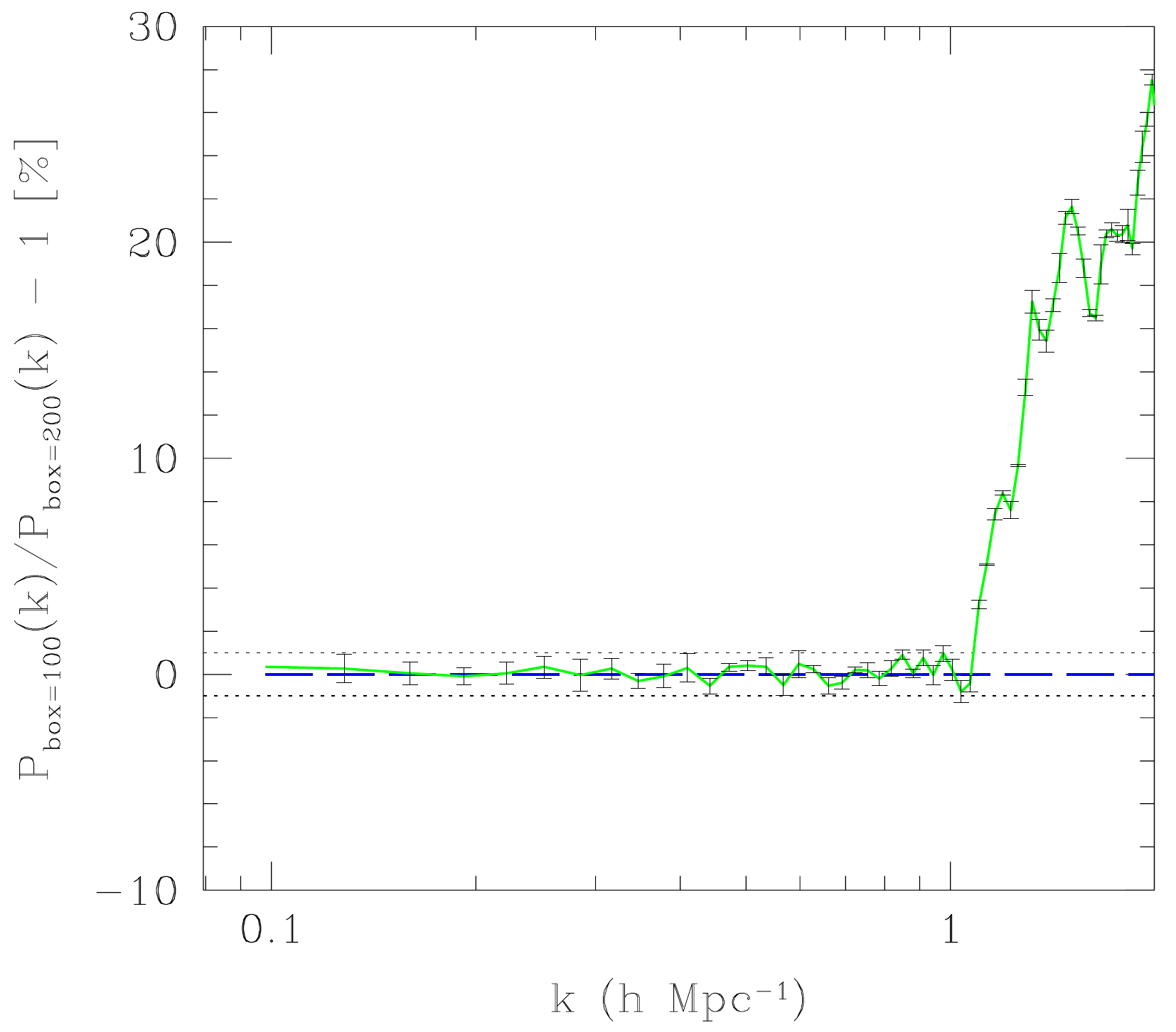}
        \caption{
Same as Fig.~\ref{fig:ps_box} expressed as fractional suppression of the matter power spectrum at $z\!=\!0$ as a function of the box size. Spectrum from $100\, h^{-1}\textrm{Mpc $-$ solid (green)}$ and $200\, h^{-1}\textrm{Mpc $-$ dashed (blue)}$ box agree at $1$ per cent level for $k\!\ltwid\!1\, h \textrm{Mpc}^{-1}$.
        }
    \label{fig:supp_box}
\end{figure}

Fig.~\ref{fig:undersampling1} shows the matter power spectrum at $z=0$ when $N_{\rm cdm}\!=\!64^3,\,128^3,\, 256^3$ and $512^3$ particles are used to sample the ICs ($\Omega_\nu=0$ for all four cases). Beyond the Nyquist wavenumbers, represented by vertical lines [$64^3$ -- solid (red), $128^3$ -- long dash--dotted (green), $256^3$ -- dashed (blue) and $512^3$ -- long-dashed (cyan)], the power spectra become increasingly inaccurate due to particle shot noise contribution. Fig.~\ref{fig:undersampling2} shows the fractional suppression of the matter power spectrum at $z=0$. For $k\!\ltwid\!0.6\, h \textrm{Mpc}^{-1}$, the error due to undersampling the ICs is $\ltwid\!5$ per cent for the $64^3$ run, $\ltwid\!0.5$ per cent for the $128^3$ run and negligibly small for the $256^3$ run. To keep the undersampling error at $k\!=\!0.6\, h \textrm{Mpc}^{-1}$ below $0.5$ per cent, we narrowed down to a combination of $N_{\rm cdm}\!=\!256^3$, $N_{\rm gas}\!=\!512^3$ in a $200\, h^{-1}\textrm{Mpc}$ box to investigate the effect of massive neutrinos on the matter power spectrum in the semi-non-linear regime $(0.1\leq k \leq 0.6\, h \textrm{Mpc}^{-1})$. Finally, we checked the smallest scales that are accurately resolved by the $200\, h^{-1}\textrm{Mpc}$ box. Towards this, we ran eight simulations in a $100\, h^{-1}\textrm{Mpc}$ box with $N_{\rm cdm}\!=\!256^3$, $N_{\rm gas}\!=\!512^3$. In Fig.~\ref{fig:ps_box}, we plot the power spectrum from $100$ and $200\, h^{-1}\textrm{Mpc}$ boxes. The matter power spectrum from $100\, h^{-1}\textrm{Mpc}$ box simulations begins to show excess power for $k\!\gtwid\!1\, h \textrm{Mpc}^{-1}$. The non-linear evolution of perturbations on scales $k\!\gtwid\!1\, h \textrm{Mpc}^{-1}$ is missed in the $200\, h^{-1}\textrm{Mpc}$ box simulations. The spectrum from $200\, h^{-1}\textrm{Mpc}$ box simulations show convergence at $1$ per cent level for $k\!\ltwid\!1\, h \textrm{Mpc}^{-1}$ (Fig.~\ref{fig:supp_box}).


\section{Impact of massive neutrinos}

The contribution of massive neutrinos to the present-day critical energy density is given by:
\be
\Omega_\nu=\frac{\Sigma m_\nu}{94.22 h^2}
\ee
where $\Sigma m_\nu$ is the sum of the masses of all neutrino eigentstates. In this section we consider four neutrino models: $\Omega_\nu\,=\,0,\,0.01,\,0.02$ and $0.04$ corresponding to $\Sigma m_\nu=0,\,0.475,\,0.95$ and $1.9\,{\rm eV}$, respectively. We assume three degenerate neutrino eigentstates, so that $m_\nu=\Sigma m_\nu/3$.\\

\begin{figure}
  \begin{center}
        \includegraphics[width= \columnwidth,angle=0,scale=0.945]{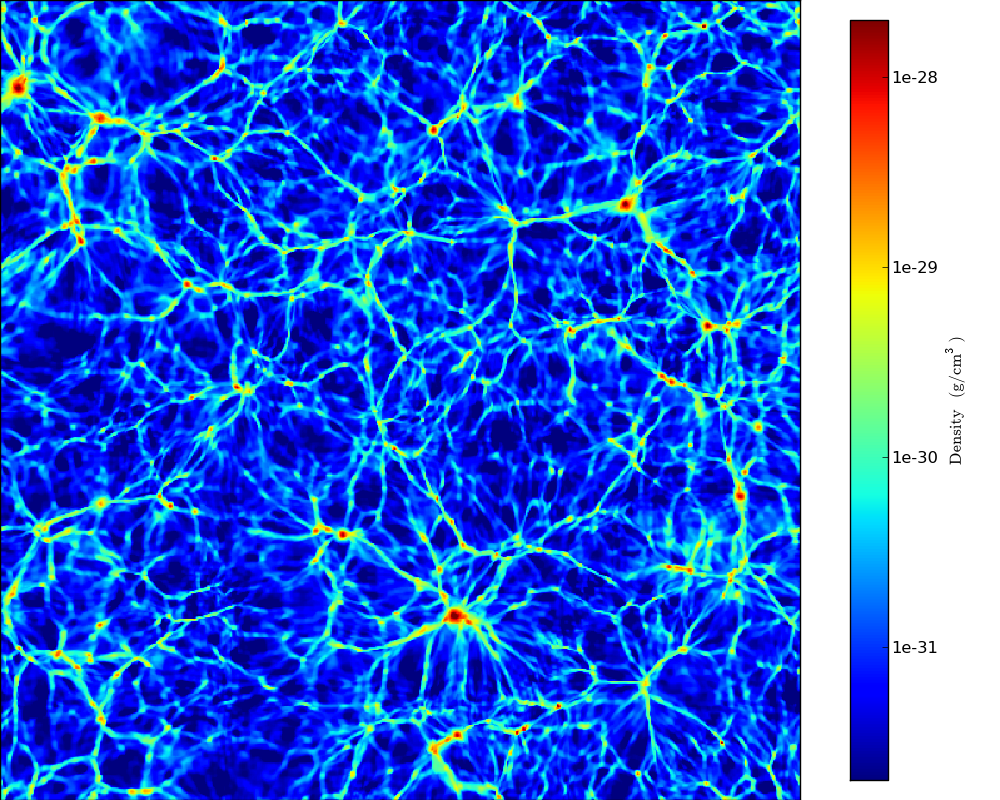}\vspace{1mm}
        \includegraphics[width= \columnwidth,angle=0,scale=0.945]{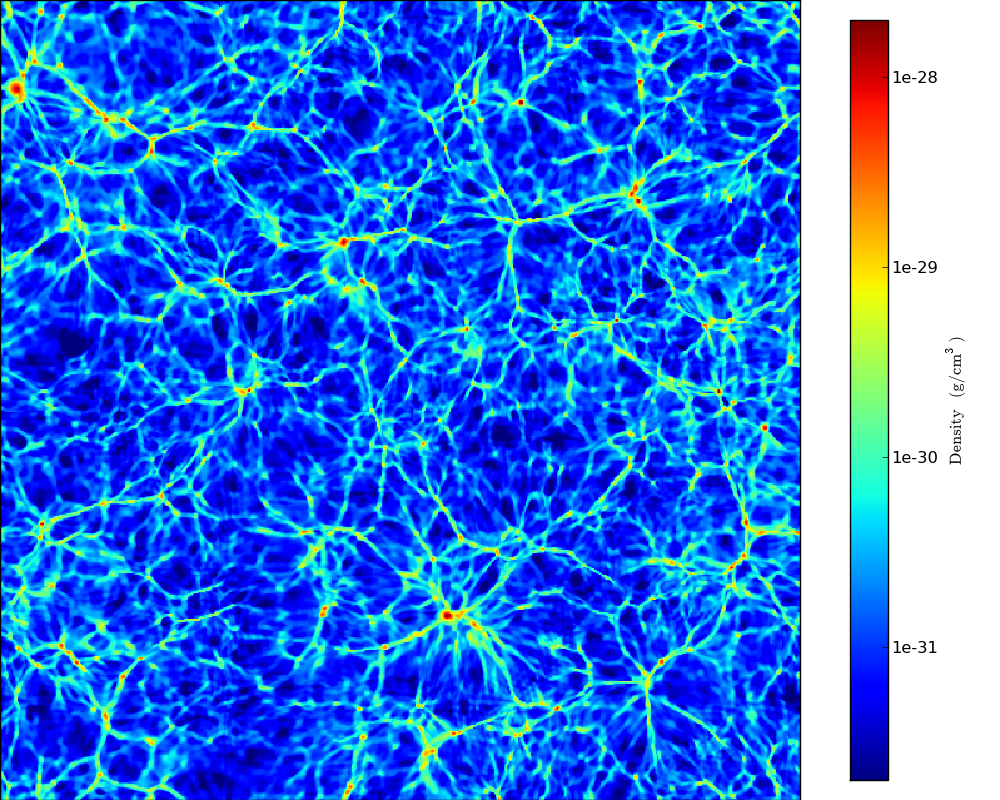}\vspace{1mm}
        \includegraphics[width= \columnwidth,angle=0,scale=0.945]{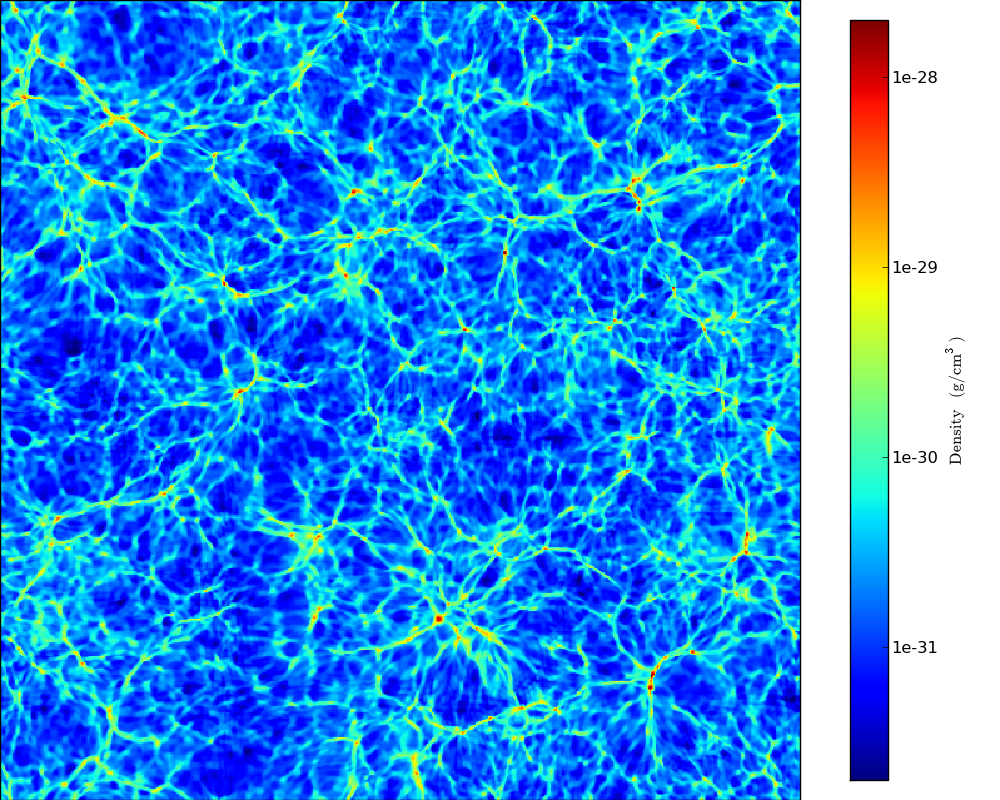}
  \end{center}
        \caption{
Slices of baryon density distribution. All slices are $200\, h^{-1}\textrm{Mpc}$ wide and show the baryonic mass averaged over the volume of a grid cell. Each grid cell is $\twid\!391\, h^{-1}\textrm{kpc}$. The top panel shows a simulation without neutrinos. The middle and the bottom panels are taken from simulations with $\Omega_\nu=0.02\,(\Sigma m_\nu=0.95\,{\rm eV})$ and $\Omega_\nu=0.04\,(\Sigma m_\nu=1.9\,{\rm eV})$. The baryon density fields in the middle and the bottom panels are less evolved relative to the no-neutrino (top panel) case. The simulations were run with $N_{\rm cdm}\!=\!256^3$, $N_{\rm gas}\!=\!512^3$. The density projections were made using {\sc yt}: an analysis and visualization tool \citep{SciPyProceedings_46}.
        }
    \label{fig:0eV_Density}
\end{figure}

In Fig.~\ref{fig:0eV_Density} we show slices of the baryon density field at $z=0$ extracted from $200\, h^{-1}\textrm{Mpc}$ box with $N_{\rm cdm}\!=\!256^3$, $N_{\rm gas}\!=\!512^3$. The top panel is from a simulation without neutrinos, the middle and the bottom panels correspond to simulations with $\Omega_\nu=0.02$ and $0.04$ respectively. All slices are $200\, h^{-1}\textrm{Mpc}$ wide. The slices show the baryonic mass averaged over the volume of a grid cell. Each grid cell in our simulations is $\twid\!391\, h^{-1}\textrm{kpc}$.

As neutrinos become more massive, the suppression in the growth of density perturbations becomes clear by the relatively diffused density filaments. The baryon density fields in the middle and the bottom panels are less evolved relative to the massless neutrino (top panel) case. The gravitational potential wells are much deeper in the top panel. This is evident from the voids (dark blue regions) which are more underdense in the top panel compared to the voids in the lower panels. To quantify the difference between simulations with and without massive neutrinos, we measure the total matter power spectrum by converting the positions of the CDM and gas particles into $512^3$-point grids of densities using a Cloud-In-Cell (CIC) interpolation scheme. We do not compensate for the smoothing effect introduced by the CIC filtering since the smoothing affects scales that are close to the Nyquist wavenumber which for our choice of parameters ($N_{\rm gas}\!=\!512^3$, Box=$200\, h^{-1}\textrm{Mpc}$) is $k_{\rm Ny}=8.04\, h\textrm{Mpc}^{-1}$, while the semi-non-linear modes of interest are $0.1\!\ltwid\!k\!\ltwid\!0.6\, h \textrm{Mpc}^{-1}$. The density fields are fast Fourier transformed to calculate $P_{\rm b}^{\rm NL}(k)$ and $P_{\rm c}^{\rm NL}(k)$ -- the non-linear power spectrum for baryons and CDM respectively. We then construct $P^{\rm NL}_{\rm m}(k)$ at $z=0$ using equations (8) and (9). To suppress sampling variance of the estimated $P(k)$, we take the average $P(k)$ from eight simulations.

\begin{figure}
     \includegraphics[width= \columnwidth]{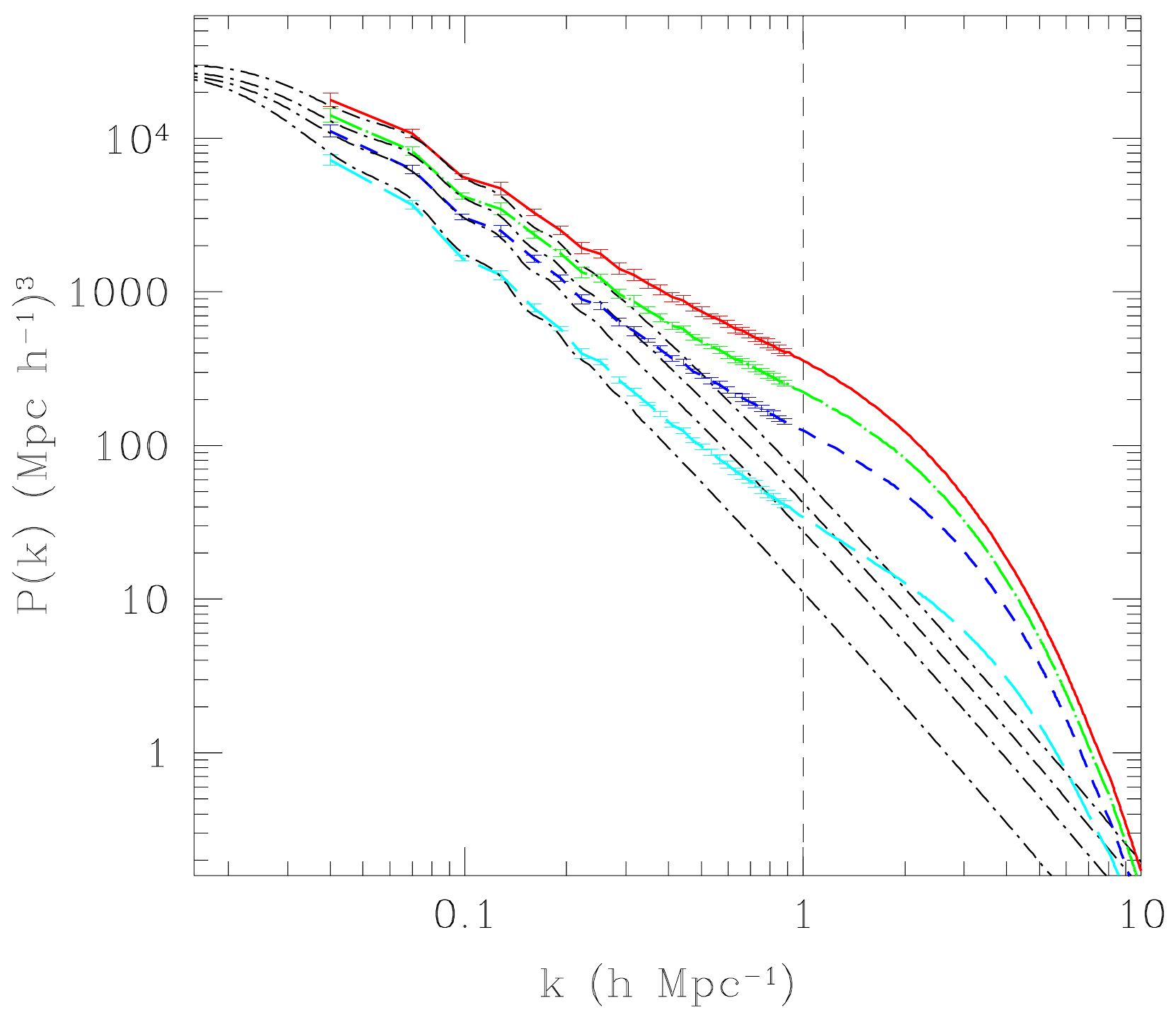}
        \caption{
Matter power spectrum at $z=0$ from simulations and linear theory (dot--dashed lines) as a function of neutrino mass. The four neutrino models are: $\Omega_\nu\!=\!0\,(\Sigma m_\nu=0\,{\rm eV})$ -- solid (red), $\Omega_\nu\!=\!0.01\,(\Sigma m_\nu=0.475\,{\rm eV})$ -- long dash--dotted (green), $\Omega_\nu\!=\!0.02\,(\Sigma m_\nu=0.95\,{\rm eV})$ -- dashed (blue) and $\Omega_\nu\!=\!0.04\,(\Sigma m_\nu=1.9\,{\rm eV})$ -- long-dashed (cyan). The vertical dashed line is the maximum wavenumber up to which the power spectra from $200\, h^{-1}\textrm{Mpc}$ box simulations are valid at $1$ per cent level.
        }
    \label{fig:ps1}
\end{figure}

\begin{figure}
     \includegraphics[width= \columnwidth]{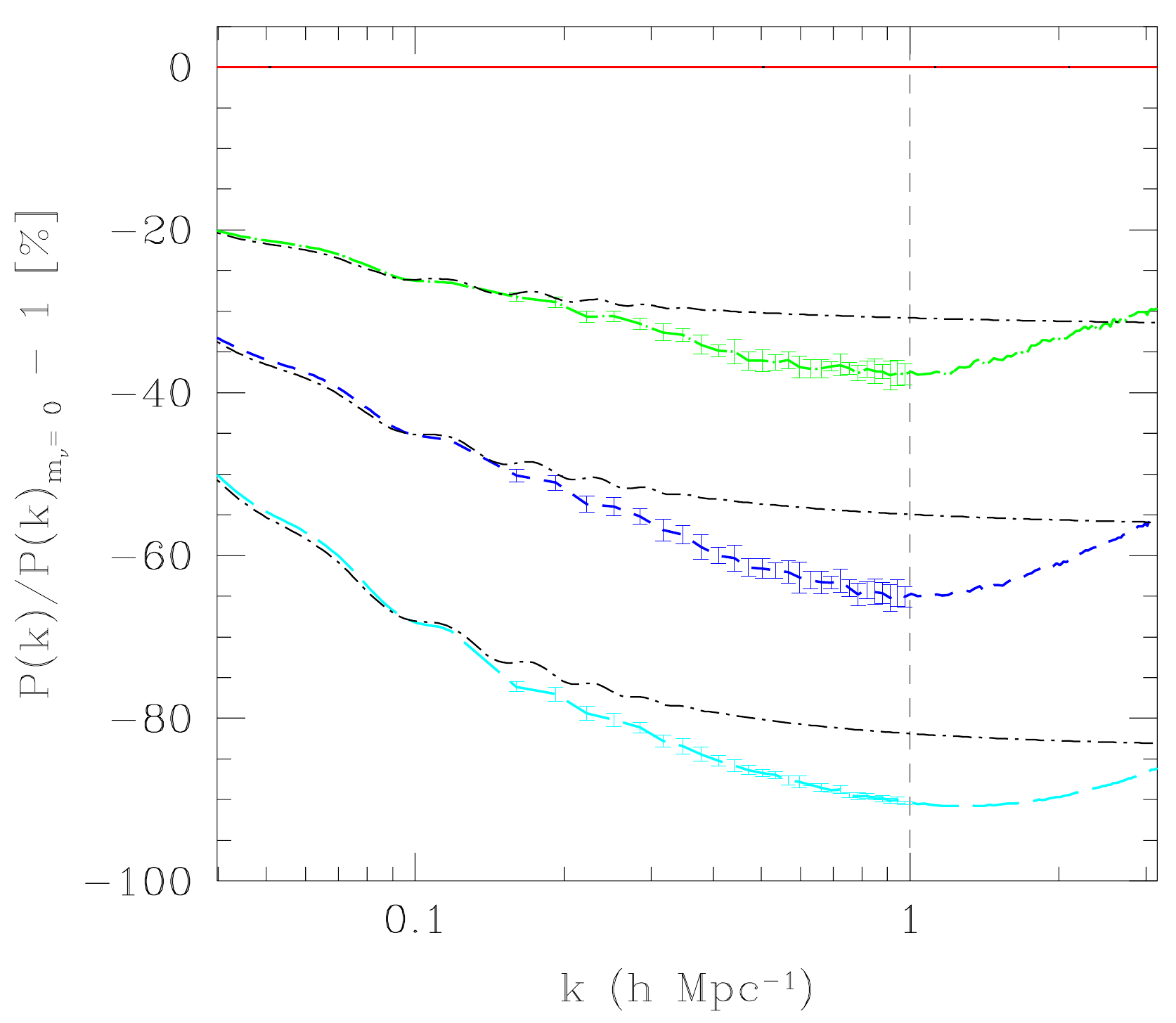}
        \caption{
Fractional difference between the matter power spectra with and without massive neutrinos at $z=0$, from the simulations and the linear theory predictions (dot--dashed lines). The four neutrino models are: $\Omega_\nu\!=\!0\,(\Sigma m_\nu=0\,{\rm eV})$ -- solid (red), $\Omega_\nu\!=\!0.01\,(\Sigma m_\nu=0.475\,{\rm eV})$ -- long dash--dotted (green), $\Omega_\nu\!=\!0.02\,(\Sigma m_\nu=0.95\,{\rm eV})$ -- dashed (blue) and $\Omega_\nu\!=\!0.04\,(\Sigma m_\nu=1.9\,{\rm eV})$ -- long-dashed (cyan). The error bars correspond to eight simulations with different seeds for the ICs.
        }
    \label{fig:ps_supp1}
\end{figure}

Fig.~\ref{fig:ps1} shows the matter power spectrum at $z\!=\!0$ from simulations and linear theory (dot--dashed lines) as a function of neutrino mass for the four neutrino models: $\Omega_\nu\!=\!0\,(\Sigma m_\nu=0\,{\rm eV})$ -- solid (red), $\Omega_\nu\!=\!0.01\,(\Sigma m_\nu=0.475\,{\rm eV})$ -- long dash-dotted (green), $\Omega_\nu\!=\!0.02\,(\Sigma m_\nu=0.95\,{\rm eV})$ -- dashed (blue) and $\Omega_\nu\!=\!0.04\,(\Sigma m_\nu=1.9\,{\rm eV})$ -- long-dashed (cyan). The simulation spectra are significantly above the linear theory predictions at high k. The linear theory predictions break down for $k\!\gtwid\!0.1\,h\textrm{Mpc}^{-1}$ ($\lambda\!\ltwid\!60\,h^{-1}\textrm{Mpc}$). Also, as the total neutrino mass is increased (keeping the number of degenerate neutrino eigentstates fixed at three), the matter power spectrum is further suppressed. Since neutrino eigentstates with higher mass constitute a larger fraction of the total energy density, they are more effective in damping small-scale power than low mass neutrinos.

In Fig.~\ref{fig:ps_supp1} we plot the fractional difference between the matter power spectra with and without massive neutrinos, from the simulations as well as the linear theory predictions. The linetypes for the spectra are the same as in Fig.~\ref{fig:ps1}. The linear theory predicts a nearly scale-independent suppression for $k\!\gtwid\!0.2\, h\textrm{Mpc}^{-1}$. On the other other hand, the non-linear power spectra from the simulations show an enhanced suppression for $k\!\gtwid\!0.1\, h\textrm{Mpc}^{-1}$. At $k\!\twid\!1\, h\textrm{Mpc}^{-1}$, the non-linear spectra are $\twid\!10$ per cent more suppressed compared to the corresponding linear spectra.

\section{Resolving neutrino mass hierarchy from simulations}

The mass splittings of $|\Delta m^{2}_{32}|=(2.43\pm0.13)\times10^{-3}\,{\rm eV}^2$ and $\Delta m^{2}_{21}=(7.59\pm0.21)\times10^{-5}\,{\rm eV}^2$ \citep{MINOS,KamLAND} allow for two possible neutrino mass hierarchies: normal ($m_3>m_2>m_1$) and inverted ($m_2>m_1>m_3$). For $\Sigma m_\nu>0.4-0.5\,{\rm eV}$, all neutrino eigentstates are essentialy degenerate, the mass of each eigentstate being $m_\nu\approx\Sigma m_\nu/3$. However, for smaller $\Sigma m_\nu$, the individual eigentstate masses differ significantly in the normal and inverted hierarchies. The free-streaming comoving wave number, $k_{\rm nr}$, is a function of the mass of each neutrino eigentstate (see equations 4 and 5). As the mass  is increased, it becomes non-relativistic earlier and the free-streaming scale gets shorter. The mass dependence of $k_{\rm nr}$ means that the matter power spectrum is modified differently for eigentstates with different masses. This makes the matter power spectrum a powerful tool to distinguish between the normal and inverted hierarchies. In this section we discuss the precision levels above which the power spectrum from future galaxy surveys should be able to resolve between the two mass hierarchies.
 
\begin{figure}
     \includegraphics[width= \columnwidth]{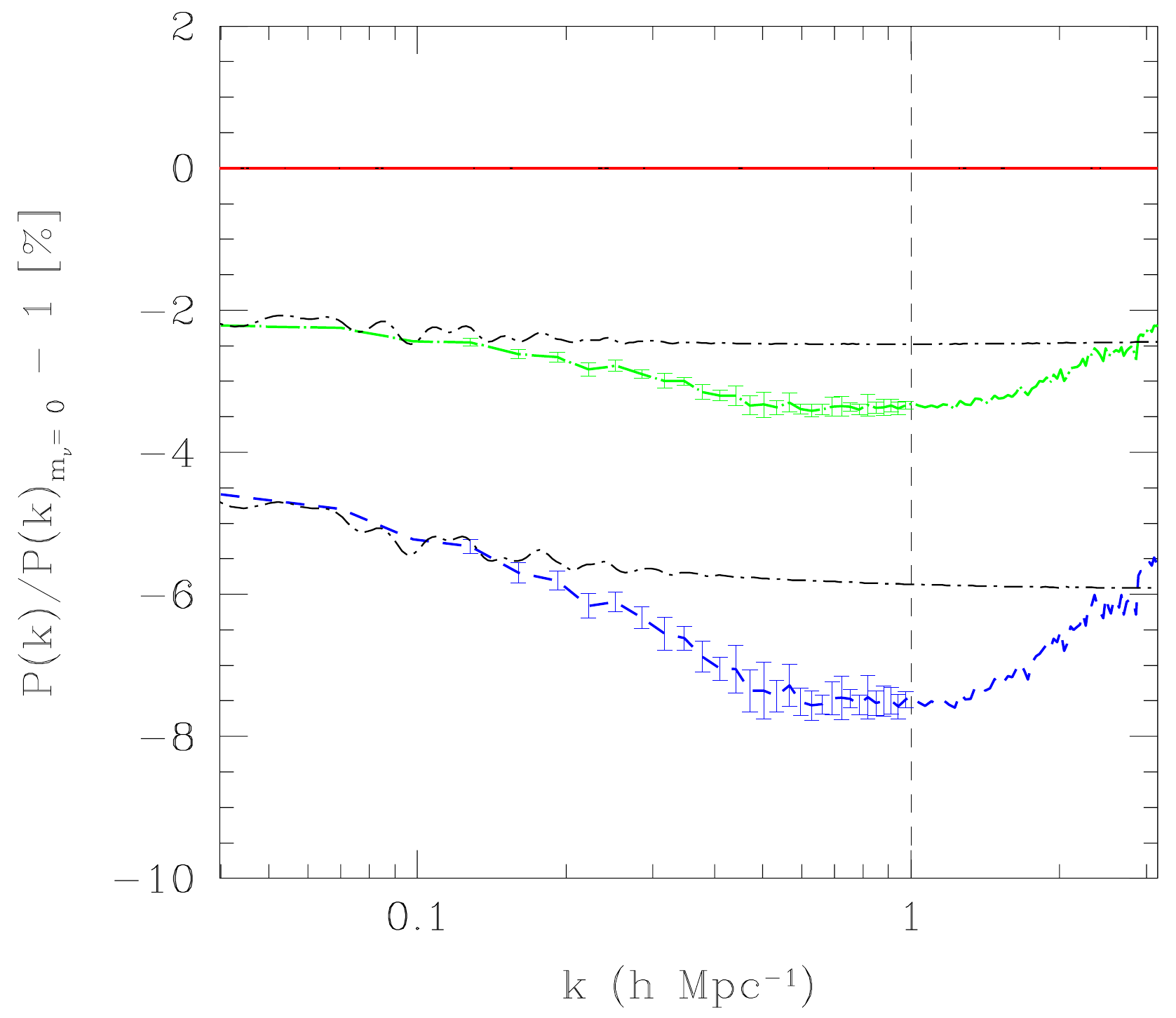}
        \caption{
Same as Fig.~\ref{fig:ps_supp1}, but for neutrino models with much lower neutrino mass: $\Omega_\nu\!=\!0.001\,(\Sigma m_\nu=0.05\,{\rm eV})$ -- long dash--dotted (green) and $\Omega_\nu\!=\!0.002\,(\Sigma m_\nu=0.1\,{\rm eV})$ -- dashed (blue).
        }
    \label{fig:ps_supp2}
\end{figure}

The mass splittings of $|\Delta m^{2}_{32}|=(2.43\pm0.13)\times10^{-3}\,{\rm eV}^2$ and $\Delta m^{2}_{21}=(7.59\pm0.21)\times10^{-5}\,{\rm eV}^2$ imply that the lower bounds on the total neutrino mass are $\Sigma m_\nu=0.05$ and $0.1\,{\rm eV}$ for the normal and inverted mass hierarchies respectively. We performed $\it{N}$-body simulations for $\Sigma m_\nu=0.05$ and $0.1\,{\rm eV}$. For $\Sigma m_\nu=0.05\,{\rm eV}$, we assumed 1 massive and 2 massless eigentstates (mimicking the normal hierarchy). For $\Sigma m_\nu=0.1\,{\rm eV}$, we assumed 2 massive and 1 massless eigentstate (mimicking the inverted hierarchy). In Fig.~\ref{fig:ps_supp2}, we show the fractional suppression in the power spectrum for two neutrino models: $\Omega_\nu\!=\!0.001\,(\Sigma m_\nu=0.05\,{\rm eV})$ -- long dash--dotted (green) and $\Omega_\nu\!=\!0.002\,(\Sigma m_\nu=0.1\,{\rm eV})$ -- dashed (blue). The growth of structure formation is suppressed by as much as $3.5$ per cent ($7.5$ per cent) at $k\!\twid\!0.6\,h\textrm{Mpc}^{-1}$ for the two models. The measurement errors in the power spectrum from future galaxy surveys are expected to be at the $1$ per cent level. In case future surveys constrain $\Sigma m_\nu<0.1\,{\rm eV}$ with sufficient precision, that would rule out the inverted mass hierarchy. The current constraint from the 7-yr $\it{WMAP}$ data alone \citep{WMAP7b} is $\Sigma m_\nu<1.3\,{\rm eV}$ ($95$ per cent CL). At this level, it is not possible to discriminate between the normal and inverted hierarchies since all eigentstates are essentially degenerate.

Next, we consider a scenario with $\Sigma m_\nu=0.1\,{\rm eV}$, at which the difference between the normal and inverted hierarchies is most prominent. We ran $\it{N}$-body simulations in the following three ways: (i) $(N_{\rm massive}=3,$ $N_{\rm degen}=3)$ where $N_{\rm massive}$ is the number of massive eigentstates and $N_{\rm degen}$ is the degeneracy amongst the massive eigentstates. This combination corresponds to $m_\nu=\Sigma m_\nu/3=0.033\,{\rm eV}$; (ii) $(N_{\rm massive}=2,$ $N_{\rm degen}=2),$ this is the inverted hierarchy scenario with one massless  and two equally massive eigentstates ($m_\nu\twid0.05,0.05,0\,{\rm eV}$); (iii) $(N_{\rm massive}=3,$ $N_{\rm degen}=2),$ this is the normal hierarchy scenario with three massive eigentstates ($m_\nu\twid0.056,0.022,0.022\,{\rm eV}$). Note that case (i) is meaningless at $\Sigma m_\nu=0.1\,{\rm eV}$ given that $|\Delta m^{2}_{32}|=(2.43\pm0.13)\times10^{-3}\,{\rm eV}^2$ and $\Delta m^{2}_{21}=(7.59\pm0.21)\times10^{-5}\,{\rm eV}^2$. We include case (i) for illustrative purposes only.

\begin{figure}
     \includegraphics[width= \columnwidth]{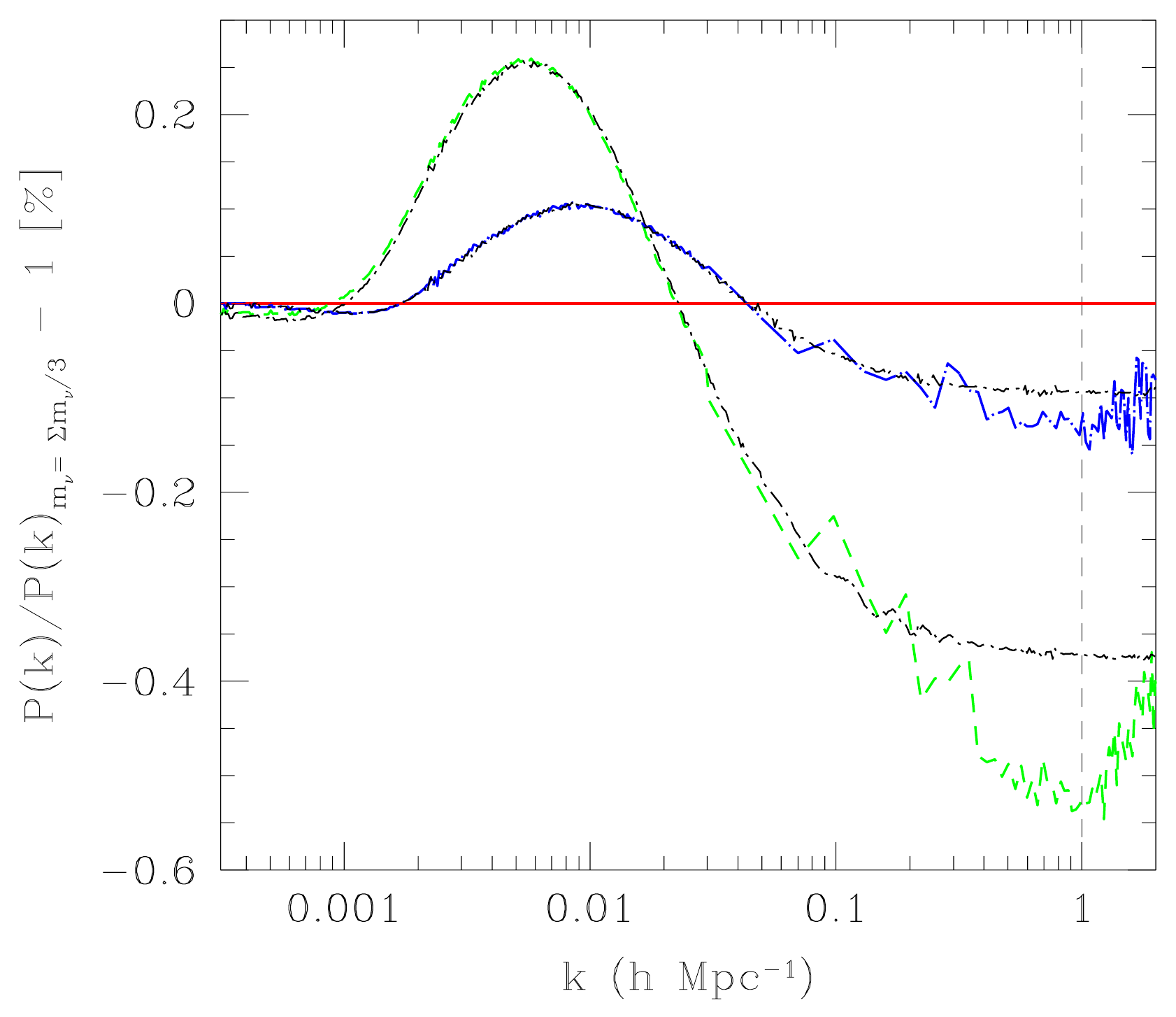}
        \caption{
Matter power spectrum for normal -- long dash--dotted line (blue) and inverted -- dashed line (green) hierarchies divided by the matter power spectrum for $m_\nu=\Sigma m_\nu/3$ -- solid line (red). The linear theory predictions are shown by dot--dashed lines. The neutrino model considered here is $\Sigma m_\nu=0\,{\rm eV}$. The individual masses for the three eigentstates are ($m_\nu\twid0.05,0.05$ and $0\,{\rm eV}$) for the inverted hierarchy and ($m_\nu\twid0.056,0.022$ and $0.022\,{\rm eV}$) for the normal hierarchy. The inverted hierarchy shows more damping of small-scale power than the normal hierarchy.
        }
    \label{fig:ps_supp3}
\end{figure}

In Fig.~\ref{fig:ps_supp3}, we plot the matter power spectrum for cases (i), (ii) and (iii) divided by the spectrum for case (i). The linear theory predictions are shown by dot--dashed lines. Since non-linearities become important only for $k\!\gtwid\!0.1\,h\textrm{Mpc}^{-1}$, we have plotted the theoretical power spectrum for $k\!<\!0.1\,h\textrm{Mpc}^{-1}$, calculated using the {\sc camb} code. The suppression from simulations is $\!\twid\!0.05-0.2$ per cent higher than the linear predictions. The inverted hierarchy - dashed line (green) shows excess power for wavenumbers $0.001\!<\!k\!<\!0.02\,h\textrm{Mpc}^{-1}$ and an enhanced suppression of $\twid\!0.5$ per cent at $k\!\twid\!1\,h\textrm{Mpc}^{-1}$ relative to case (i). This can be explained by the fact that in case (ii) $\Sigma m_\nu=0.1\,{\rm eV}$ is shared equally between two eigentstates, while in case (i) $\Sigma m_\nu=0.1\,{\rm eV}$ is shared equally between three eigentstates. Each eigentstate is more massive in case (ii), thereby making the free-streaming length shorter compared to that in case (i). Higher mass neutrinos are better at wiping out small-scale perturbations and their shorter free-streaming length implies that the spatial extent of damping is limited.

Another factor contributing to the appearance of Fig.~\ref{fig:ps_supp3} is a shift in the radiation--matter equality redshift. Higher mass neutrinos become non-relativistic at higher redshifts and start contributing to $\Omega_{\rm m}$ before low mass neutrinos do. This shifts the radiation--matter equality epoch to a higher redshift and reduces the scale corresponding to the one that entered the horizon at radiation--matter equality. The modes entering the horizon after radiation--matter equality grow linearly (as opposed to logarithmically during the radiation era) which contributes to the excess power [compare dashed (green) and solid (red) lines in Fig.~\ref{fig:ps_supp3}] for wavenumbers $0.001\!<\!k\!<\!0.02\,h\textrm{Mpc}^{-1}$. The same reasoning can be applied to the normal hierarchy -- long dash--dotted line (blue). At $\Sigma m_\nu=0.1\,{\rm eV}$, precision better than $0.5$ per cent would be needed in measuring the matter power spectrum to discriminate between the normal and inverted hierarchies. For $\Sigma m_\nu>0.2\,{\rm eV}$ all eigentstates become degenerate, this would make it extremely difficult for a future survey to resolve the two hierarchies.

\section{Comparison with recent numerical studies}

\begin{figure}
     \includegraphics[width= \columnwidth]{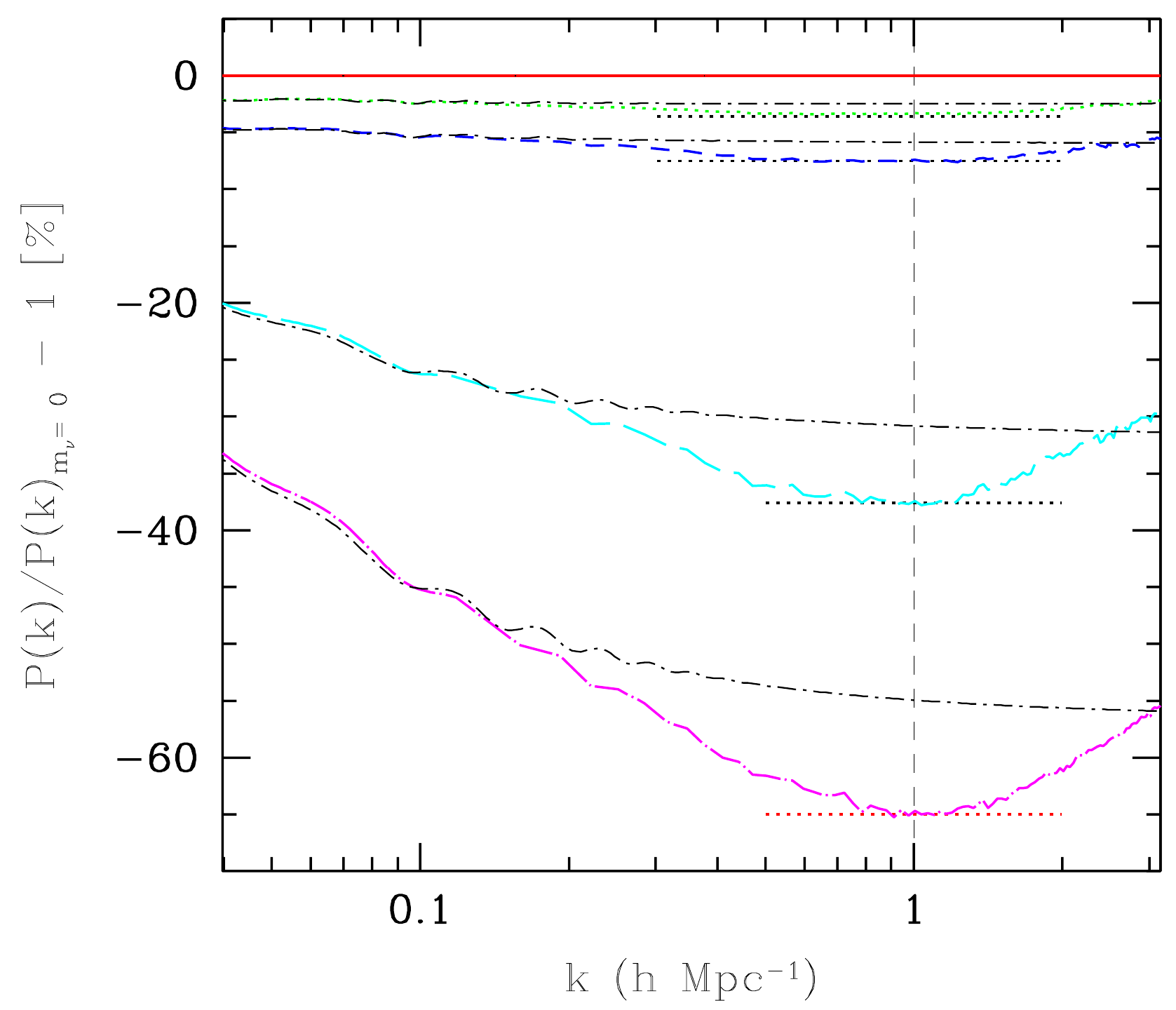}
        \caption{
Fractional difference between the matter power spectra with and without massive neutrinos at $z=0$, from the simulations and the linear theory predictions (dash--dotted lines). The four neutrino models are: $\Omega_\nu\!=\!0.001\,(\Sigma m_\nu=0.05\,{\rm eV})$ -- dotted (green), $\Omega_\nu\!=\!0.002\,(\Sigma m_\nu=0.1\,{\rm eV})$ -- dashed (blue), $\Omega_\nu\!=\!0.01\,(\Sigma m_\nu=0.475\,{\rm eV})$ -- long-dashed (cyan) and $\Omega_\nu\!=\!0.02\,(\Sigma m_\nu=0.95\,{\rm eV})$ -- long dash--dotted (magenta). The maximum relative suppression of $\Delta P/P\!\twid\!-10f_\nu$ is shown as short horizontal dotted lines. The horizontal (red) dotted line for $\Sigma m_\nu=0.95\,{\rm eV}$ is at $\Delta P/P\!\twid\!-8.6f_\nu$.
        }
    \label{fig:ps_supp4}
\end{figure}

In this section we compare the estimated overall suppression of the matter power spectrum due to massive neutrinos from our $\it{N}$-body simulations with the results obtained by \cite{BraHan08} and \cite{VieHaeSpr10}. In linear theory, the suppression of the matter power spectrum amplitude is approximately given by $\Delta P/P\!\twid\!-8f_\nu$ \citep*{HuEisTeg98}. Numerical simulations, however, show that the neutrino suppression is enhanced in the non-linear regime $(k\!\gtwid\!0.1\,h\textrm{Mpc}^{-1})$. In Fig.~\ref{fig:ps_supp4} we plot the fractional difference between the matter power spectra with and without massive neutrinos at $z=0$, from the simulations as well as the linear theory predictions (dash--dotted lines) for four neutrino models: $\Omega_\nu\!=\!0.001\,(\Sigma m_\nu=0.05\,{\rm eV})$ -- dotted (green), $\Omega_\nu\!=\!0.002\,(\Sigma m_\nu=0.1\,{\rm eV})$ -- dashed (blue), $\Omega_\nu\!=\!0.01\,(\Sigma m_\nu=0.475\,{\rm eV})$ -- long-dashed (cyan) and $\Omega_\nu\!=\!0.02\,(\Sigma m_\nu=0.95\,{\rm eV})$ -- long dash--dotted (magenta). We found a maximum non-linear suppression of $\Delta P/P\!\twid\!-10f_\nu$ for neutrino masses $\Sigma m_\nu=0.05, 0.1, 0.475\,{\rm eV}$. Although we ran our simulations with a slightly different set of cosmological parameters, \cite{BraHan08} measured $\Delta P/P\!\twid\!-9.8f_\nu$ for $\Sigma m_\nu\leq0.6\,{\rm eV}$ while \cite{VieHaeSpr10} reported $\Delta P/P\!\twid\!-9.5f_\nu$ at $z=0$. For $\Sigma m_\nu=0.95\,{\rm eV}$, we get $\Delta P/P\!\twid\!-8.6f_\nu$ while \cite{VieHaeSpr10} reported $\Delta P/P\!\twid\!-8f_\nu$ for $\Sigma m_\nu=1.2\,{\rm eV}$. The scale at which the suppression turns over, $k_{\rm nr}$, moves from $k_{\rm nr}\!\twid\!0.6-0.7\,h\textrm{Mpc}^{-1}$ for $\Sigma m_\nu=0.05\,{\rm eV}$ to $k_{\rm nr}\!\twid\!1\,h\textrm{Mpc}^{-1}$ for $\Sigma m_\nu=0.95\,{\rm eV}$. The turnover may be related to the non-linear collapse of structures as discussed in \cite{BraHan08} who reported $k_{\rm nr}\!\twid\!1\,h\textrm{Mpc}^{-1}$.

\section{Matter Power Spectrum Error Estimates}

In our $\it{N}$-body simulations, we have implemented neutrinos in the ICs only. Neutrino-weighted CDM and baryon transfer functions from {\sc camb} were used to generate the ICs for CDM particles and baryons. To construct $P^{\rm NL}_{\rm m}(k)$ at $z=0$, we used equation (8), where $P^{\rm NL}_{\rm cb}$ at $z=0$ from $\it{N}$-body simulations was combined with $P_\nu^{\rm L}$ at $z=0$ as solved by the {\sc camb} code. This methodology introduces errors in the estimated matter power spectrum for two reasons: (i) the linear neutrino perturbations were taken into account only at the initial ($z_{\rm i}=20$) and the final ($z=0$) redshifts. There is no feedback from the neutrinos on to the CDM component in our $\it{N}$-body simulations. (ii) the non-linear evolution of neutrino perturbations was not accounted for in our $\it{N}$-body simulations. While the extent of non-linear neutrino corrections to the matter power spectrum is still being studied, we use \cite{BraHan08} and \cite{BraHan09} to estimate the errors in our $\it{N}$-body spectra. \cite{BraHan09} describe the linear neutrino density on a grid and evolve this density forward in time using linear theory. The neutrino contribution is added to the CDM component when calculating the gravitational forces. Thus, the linear neutrino component is accounted for recursively over the redshift range over which the matter power spectrum is to be evolved. \cite{BraHan08} (their fig. 7, left panel) show that the matter power spectrum is underevolved by $\!\twid\!3$ per cent for $\Sigma m_\nu\leq0.6\,{\rm eV}$ on scales $k\geq0.2\,h\textrm{Mpc}^{-1}$ when the neutrino grid is neglected. Accordingly, our matter power spectrum estimates are expected to be underevolved by roughly $\!\ltwid\!5,3$ and $0.1$ per cent for $\Sigma m_\nu=0.95, 0.475$ and $0.1\,{\rm eV}$, respectively, for $k\!\gtwid\!0.2\,h\textrm{Mpc}^{-1}$ at $z=0$. Fig. 1 in \cite{BraHan09} shows that the power is further suppressed by $\!\twid\!5$ per cent for $\Sigma m_\nu\leq1.2\,{\rm eV}$ at $k\approx0.2-0.3\,h\textrm{Mpc}^{-1}$ when the neutrino non-linearities are neglected. Overall, we estimate our $\it{N}$-body spectrum errors to be $\!\ltwid\!10,4$ and $0.1$ per cent for $\Sigma m_\nu=0.95, 0.475$ and $0.1\,{\rm eV}$, respectively, for $k\!\gtwid\!0.2\,h\textrm{Mpc}^{-1}$ at $z=0$.

\section{Discussion and conclusions}

In this paper we simulated the matter power spectrum at $z=0$ in order to study how massive neutrinos impact structure formation. The most important factors in obtaining an accurate power spectrum are (i) the Nyquist wavenumber, which depends on the simulation box size and the number of particles and (ii) the force resolution, which depends on the size of the root grid. Above the Nyquist wavenumber, the power spectrum is dominated by shot noise. For the semi-non-linear modes $(0.1\!\ltwid\!k\!\ltwid\!0.6\, h \textrm{Mpc}^{-1})$, we found that $N_{\rm cdm}\!=\!256^3$ in a $200\,h^{-1}\textrm{Mpc}$ box is enough to keep the sampling errors under $0.5$ per cent. We used a root grid of $N_{\rm gas}\!=\!512^3$, which is twice as fine as $N_{\rm cdm}$, to accurately calculate the gravitational forces down to the scale of the mean interparticle spacing. We also found that the non-linear evolution of perturbations are accurate to within $1$ per cent level only for the scales $k\!\ltwid\!1\,h\textrm{Mpc}^{-1}$ when using $200\,h^{-1}\textrm{Mpc}$ box. Probing smaller scales with higher precision requires a smaller simulation box or a finer root grid.

We have presented a suite of $\it{N}$-body simulations showing the effect of massive neutrinos in the range $\Omega_\nu=0.001-0.04$ ($\Sigma m_\nu=0.05-1.9\,{\rm eV}$) on the distribution of matter. Massive neutrinos smooth the neutrino density field on sub-free-streaming scales. This makes the gravitational potential wells shallower than their counterparts in a pure $\Lambda CDM$ universe, leading to a suppressed growth of structure formation. The power is suppressed by as much as $3.5-90$ per cent at $k\!\twid\!0.6\,h\textrm{Mpc}^{-1}$ for $\Sigma m_\nu=0.05-1.9\,{\rm eV}$ respectively. In our simulations, we include neutrinos as neutrino-weighted CDM and baryon transfer functions at the starting redshift, $z_{\rm i}=20$. We have neglected the non-linear neutrino corrections to the matter power spectrum which may be as high as $1.25$ per cent for $\Sigma m_\nu=0.6\,{\rm eV}$ and $5$ per cent for $\Sigma m_\nu=1.2\,{\rm eV}$ as measured by \cite{BraHan09}. Although direct comparison of our $\it{N}$-body results with those from \cite{BraHan09} was not possible since we ran our simulations with a slightly different set of cosmological parameters and $\Sigma m_\nu$, nevertheless, we expect our $\it{N}$-body power spectra to be in error by $\!\ltwid\!10,4$ and $0.1$ per cent for $\Sigma m_\nu=0.95, 0.475$ and $0.1\,{\rm eV}$, respectively, for $k\!\gtwid\!0.2\,h\textrm{Mpc}^{-1}$ at $z=0$. We found an overall suppression of power from our simulations at $z=0$ to be $\Delta P/P\!\twid\!-10f_\nu$ for $\Sigma m_\nu\leq0.5\,{\rm eV}$ which is slightly higher than the results of \cite{BraHan08} and \cite{VieHaeSpr10} who reported $\Delta P/P\!\twid\!-9.8f_\nu$ and $\Delta P/P\!\twid\!-9.5f_\nu$ respectively for $\Sigma m_\nu\leq0.6\,{\rm eV}$.

As part of the Sloan Digital Sky Survey-III, the Baryon Oscillation Spectroscopic Survey (BOSS; \citealt{BOSS}) is expected to measure the power spectrum with precisions at which $\Omega_\nu\!\twid\!0.01\,(\Sigma m_\nu\twid\!0.475\,{\rm eV})$ could be ruled out. This would significantly improve the current 7-yr {\it WMAP} data alone constraint of $\Sigma m_\nu<1.3\,{\rm eV}$. If $\Sigma m_\nu$ constraints from cosmology get as low as $0.1-0.2\,{\rm eV}$, it will open up a possibility to resolve the normal and inverted mass hierarchies, though the matter power spectrum would need to be determined with precision levels well below $0.5$ per cent.

\section{Acknowledgments}

Computations described in this work were performed using the {\sc enzo} code developed by the Laboratory for Computational Astrophysics at the University of California in San Diego (http://lca.ucsd.edu). We thank the users of {\sc yt} (python-based package for analysing {\sc enzo} datasets) and {\sc enzo} for useful discussions and guidance towards running and analyzing simulations. We thank the referee for a useful and constructive report. This work was supported by the National Science Foundation through TeraGrid resources provided by the NCSA and by a grant from the Research Corporation. HAF has been supported in part by an NSF grant AST-0807326, by the University of Kansas General Research Fund (KUGRF) and acknowledges the hospitality of University College, London and Imperial College in the UK and the Institut d'Astrophysique de Paris, France.
\bibliographystyle{mn2e}
\bibliography{neutrino}

\end{document}